\documentstyle[preprint,aps]{revtex} 

\input psfig
\newcommand{\psplot}[1]{ \hbox to\textwidth{\hfill
            \psfig{figure=#1.ps,width=\textwidth}\hfill} }
\psfull    
\input psfig
\newcommand{\psplotp}[1]{ \hbox to\textwidth{\hfill
            \psfig{figure=#1.ps,width=\textwidth}\hfill} }

\begin{document}


\tighten 
\preprint{\vbox{\hbox{CLNS 97/1461\ \ \ \ \  \hfill}
                \hbox{CLEO 97--2\ \   \hfill}
}}
 
\title{
Study of the $B^0$ Semileptonic Decay Spectrum 
at the $\Upsilon$(4S) Resonance
}

\date{\today}
\maketitle

\begin{abstract}
We have made a first measurement of the lepton momentum spectrum
in a sample of events enriched in neutral
$B$'s through a
partial reconstruction of $B^0\rightarrow D^{*-}\ell^+\nu$.
This spectrum, measured with
2.38~fb$^{-1}$ of data collected at the
$\Upsilon$(4S) resonance  by the CLEO~II detector, 
is compared directly to the inclusive lepton spectrum from all
$\Upsilon$(4S) events in the same data set.
These two spectra are consistent with having the same shape above
1.5~GeV/c.
From the two spectra and two other CLEO measurements,
we obtain the $B^0$ and $B^+$ semileptonic branching 
fractions, $b_0$ and $b_+$, their ratio, and the production
ratio $f_{+-}/f_{00}$ of $B^+$ and $B^0$ pairs at the
$\Upsilon$(4S).
We report
$b_+/b_0=0.950{^{+0.117}_{-0.080}} \pm 0.091$, 
$b_0=(10.78\pm 0.60\pm 0.69)\%$,
and $b_+=(10.25\pm 0.57\pm 0.65)\%$.
$b_+/b_0$ is equivalent to the ratio of charged
to neutral $B$ lifetimes, $\tau_+/\tau_0$.  

\end{abstract}
\pacs{14.40.Nd, 13.20.He}

\begin{center}
M.~Artuso,$^{1}$ A.~Efimov,$^{1}$ F.~Frasconi,$^{1}$
M.~Gao,$^{1}$ M.~Goldberg,$^{1}$ D.~He,$^{1}$ S.~Kopp,$^{1}$
G.~C.~Moneti,$^{1}$ R.~Mountain,$^{1}$ Y.~Mukhin,$^{1}$
% ...
\end{center}
 
\small
\begin{center}
$^{1}${Syracuse University, Syracuse, New York 13244}\\
% ...
$^{13}${University of Illinois, Champaign-Urbana, Illinois 61801}\\
$^{14}${Carleton University, Ottawa, Ontario, Canada K1S 5B6 \\
and the Institute of Particle Physics, Canada}\\
% ...
$^{20}${Ohio State University, Columbus, Ohio 43210}\\
% ...
\end{center}
\newpage
\narrowtext 

%
%

It is often argued that the individual semileptonic branching
fractions, $b_+$ and $b_0$, of $B^+$ and $B^0$
should be similar, that a heavy quark meson should
decay weakly without regard to the flavor of the light
spectator quark.
Strictly speaking, however, weak hadronic decays are not isospin invariant;
in the case of charm, the asymmetry of hadronic widths is
known to be large.
For $b$-hadrons, the resulting deviation from  unity of
the lifetime ratio $\tau(B^+)/\tau(B^0)$ is predicted to be
of order 10\%\cite{bigi}.
There has also been some speculation that differences will be manifested
in the semileptonic decay spectra \cite{bigispec}.

The method of partial reconstruction of $B^0\rightarrow D^{*-}
\ell^{+}\nu$,
where the decay ${D^{*-}}{\rightarrow}{\bar D^{0}}{\pi^{-}}$ is identified 
using only the ${\pi^{-}}$,
has been used previously by CLEO~II, to measure both the mixing parameter 
$\chi_d$\cite{mix,mssthesis}
and semileptonic branching fraction $b_0$\cite{mssthesis,b0} of $B^0$.
We present here an analysis 
to measure the leptonic spectrum of an
event sample highly enriched in $B^0$ mesons through the 
application of partial reconstruction as a {\it tag}.
This is, with minor corrections,
a measurement of the inclusive leptonic spectrum
from semileptonic $B^0$ decays.
It may be compared with the inclusive spectrum in $\Upsilon$(4S) events,
which contains a roughly equal mix of $B^0$ and  $B^+$.
By comparing the rates in untagged and tagged
events, it is possible to extract not only a value for $b_0$,
but also for $b_+$ and $\beta\equiv b_+/b_0$.
Assuming that the exclusive and inclusive semileptonic partial 
widths are equal for
$B^0$ and $B^+$, $\beta$ is then equal to the ratio of charged
to neutral $B$ lifetimes $\tau_+/\tau_0$.  
%
%

The data, collected with the CLEO~II detector\cite{detector} at 
the Cornell Electron Storage Ring (CESR),
consist of integrated luminosities 2.38~fb$^{-1}$ on the ${\Upsilon (4S)}$
resonance and 1.13~fb$^{-1}$ at a center-of-mass energy which is
lower by 60~MeV (continuum).
All events considered here must
have at least 5 well-fitted charged tracks, a
measured energy at least 0.15 times the CM energy and 
an event vertex consistent with the known interaction point.

%

$B^0$ decays are identified in the mode
$B^0{\rightarrow}{D^{*-}}
\ell^+\nu$ ($D^{*-}\rightarrow \bar D^0\pi^-$)
(inclusion of the charge conjugate
decay is implied throughout this paper),
where only the $\ell^+$ and $\pi^-$ are detected.
A large sample of $B^0$ is obtained by exploiting
the extremely low energy of the ${D^{*-}}$ decay.
The method is described fully in
Ref.~\cite{mssthesis}.
In brief, the momentum of the ${\pi^{-}}$ is scaled to obtain an approximate 
four-momentum ($\widetilde{E}_{{D^*}},\widetilde{\bf{p}}_{{D^*}}$) 
for  the $D^{*-}$:
\begin{eqnarray*}
& E_{D^*} \simeq {E_{\pi}\over E^{CM}_{\pi}}M_{D^*} \equiv \widetilde{E}_{D^*}\\
& {\bf{p}}_{D^*}\simeq
{\hat{\bf{p}}}_\pi\times{\sqrt{\widetilde{E}_{D^*}^2-M_{D^*}^2}}
\equiv\widetilde{\bf{p}}_{D^*}
\end{eqnarray*}
where $E_\pi$ is the $\pi^-$ energy, $E^{CM}_\pi=145$~MeV is the energy of the
$\pi^-$ in the $D^{*-}$ rest frame, and $M_{D^*}$ is the mass of the $D^{*-}$.
Using the approximation ${\bf p_B}\simeq 0$ allows us to calculate a
squared missing mass, 
\begin{eqnarray*}
\widetilde{M}_\nu^2\equiv (E_{\mbox{beam}}-\widetilde{E}_{{D^*}} - 
E_{\ell})^2-(\widetilde{\bf{p}}_{{D^*}}+
{\bf{p}}_{\ell})^2
\end{eqnarray*}
which approximates the squared mass of the neutrino.
Nearly all $\ell^+\pi^-$ candidates stemming from decays
$B\rightarrow D^{*-} (\pi)\ell^+\nu$ ($D^{*-}
\rightarrow \bar D^0\pi^-$), where the pion is the daughter of $D^{*-}$,
fall in the signal region, $\widetilde{M}_\nu^2>-2.0$~GeV$^2$.
Such candidates constitute our tag while all other $\ell^+\pi^-$ combinations
constitute the background.

The number of tags is counted in the entire
data set and in subsets which contain additional
leptons.
To avoid confusion, the
lepton used in the tag candidate will be referred to hereafter as the ``tag
lepton'', while the additional lepton from the remainder of the
event will be called the ``second lepton''.
We allow more than one ``second lepton'' to exist in an
event.
By sorting event subsets in 100 MeV/c increments of the momentum of
the second electron or muon, we measure their spectral distributions,
which may then be compared to the corresponding inclusive lepton 
spectra~\cite{rwang}.

Each tag lepton track must have a momentum in the range 
1.4-2.4~GeV/c and be consistent with 
originating at the interaction point. 
Electrons are identified primarily by the ratio of calorimetric energy
to momentum and specific ionization ($dE/dx$) in the drift chamber. 
Muons must penetrate at least five nuclear absorption lengths.

The pion in a tag candidate must be consistent with 
originating at the interaction point,
have charge opposite to that of the lepton, and have momentum 
less than 0.190~GeV/c.
If $dE/dx$ information is available, it must be
consistent with the pion hypothesis.

Second electrons (muons) are required to have momentum between 
0.6~GeV/c (1.5~GeV/c) and 2.5~GeV/c.
Since $B^0$ mixes with $\bar{B}^0$, we accept second leptons of
either the same or opposite sign relative to the tag lepton. 
The tracking criteria are identical to those for the tag lepton
except that the track polar angles are more restricted, 
$|\cos \theta_\ell| < 0.61$ for muons and
$|\cos \theta_\ell| < 0.707$ for electrons, where $\theta_\ell$ is
the track polar angle.
By matching the tracking and identification criteria to those used in
the determination of the inclusive spectra \cite{rwang},
we minimize systematic uncertainties in our comparison 
of the two. 
To eliminate multiply reconstructed tracks, 
a second lepton candidate is rejected
if the angle between its momentum and that of the tag lepton is
small, $\cos\theta_{\ell\ell}> 0.99$.

To obtain the net number of tags in each
event subset, we select candidates in the
signal region ($\widetilde{M}_\nu^2>-2.0$~GeV$^2$) and 
subtract backgrounds.
The nonresonant contribution is estimated using the continuum dataset, 
corrected for luminosity and energy differences.
Where a second lepton is selected, we account for
cases where this second lepton is a misidentified hadron (fake) by
repeating the above process 
selecting, in place of second leptons,
tracks not identified as leptons but passing all other
second lepton requirements.
The result is scaled by the fake rate per track at the appropriate 
momentum \cite{rwang}.
After correction for fakes,
the remaining background, from candidates which are random
combinations of leptons and pions in ${B\bar{B}}$ events,
is estimated via Monte Carlo.
Its normalization is obtained from
the sideband region, $-20<\widetilde{M}_\nu^2<-4$~GeV$^2$. 
The set of events used for this estimate is approximately
five times the size of the data set.
The bins with second lepton momentum in the range 2.1-2.5~GeV/c
are combined for the Monte
Carlo sample to improve the statistical significance
of the background estimate.
The statistical error of the Monte Carlo sample
is included with that of the data and contributes
approximately 10\% to its magnitude.
Tags that are counted more than once within the
signal region because
the soft pion curls and is multiply reconstructed
are found in Monte Carlo to occur for about 4\% of the signal.
This overcounting cancels in the ratio of tagging rates which
is taken for the final result.
The resulting overestimate of the statistics is
accounted for by increasing the statistical errors by 2\%.

The evaluation of background in the signal region 
depends on an accurate shape for the simulated background distribution 
in $\widetilde{M}_\nu^2$.  
To quantify the uncertainty on this,
we repeat the analysis using ``wrong-sign'' 
candidates, where the tag lepton and pion carry the 
same charge and no net signal is expected.
We obtain a net ``signal'' of $-277\pm 211$ events from the entire event set.
This is consistent with zero, and we take as the systematic
uncertainty the absolute value of the mean plus one standard deviation,
which corresponds to 3.0\% of raw candidates in the wrong sign signal region.
We then take the uncertainty in the right sign background
of each event subset to be the same fraction.
The resulting systematic errors for tags with a second lepton are found to be
1.4\%-2.1\% of the net signals.
We assign an overall error of 2.1\% to the second lepton spectrum.

\begin{figure}
\vspace{2.cm}
\centerline{\psfig{figure=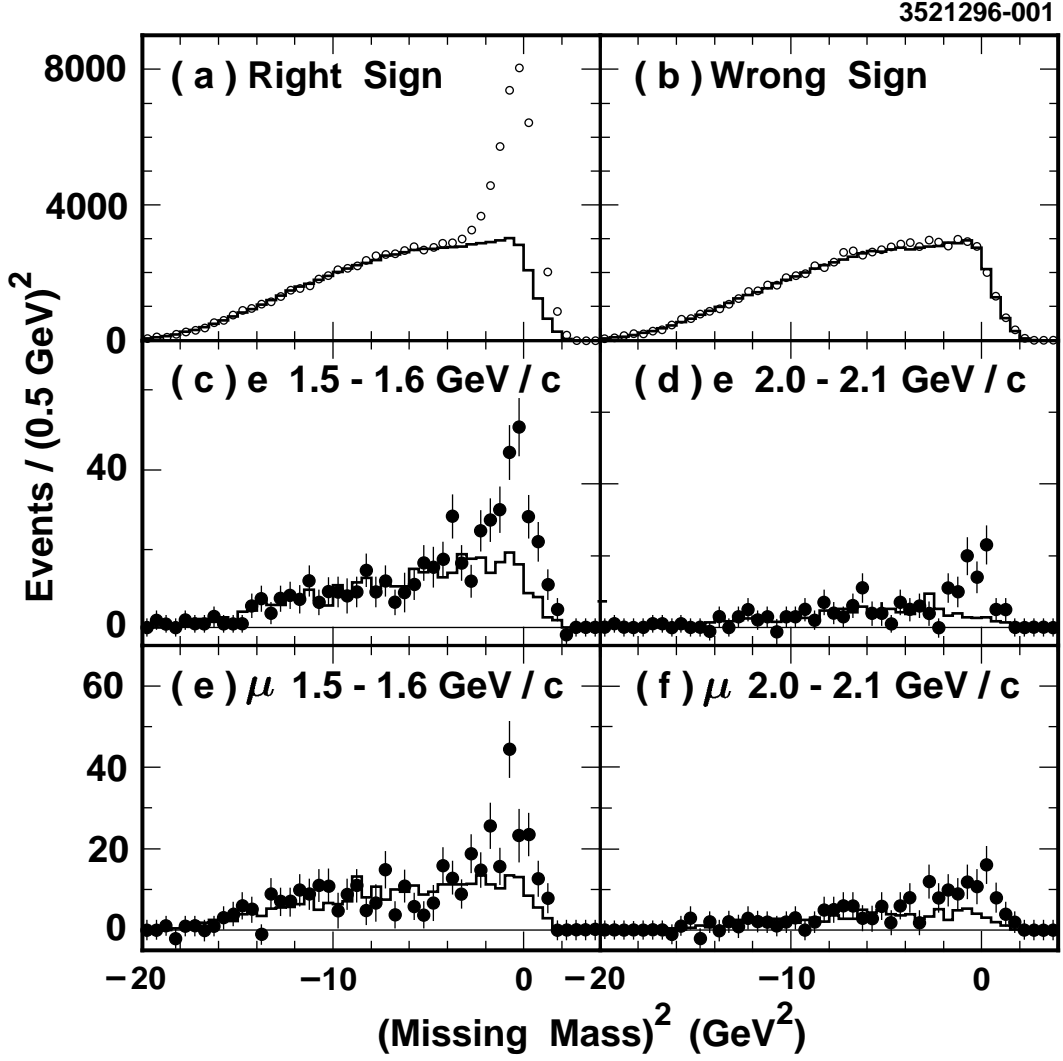}}
 \caption{
Tag candidate $\widetilde M_\nu^2$ distributions,
data (symbols) and
simulated backgrounds (solid histograms). 
Continuum has been subtracted in all cases, and
in (c)-(f), fakes have been subtracted.
(a)-(b): full event set (errors are smaller than symbols), (a)  opposite-sign
candidates, (b) same-sign.
(c)-(f): events containing a second lepton, opposite-sign candidates,
selected bins as indicated.
}
 \label{fig:tags}
\end{figure}

The net number of tags in the entire data set, $N_T$, is 
$22,880\pm 283\pm 491$.
Figure~\ref{fig:tags} shows $\widetilde{M}_\nu^2$ distributions with
estimated backgrounds for
the full data set and for subsets corresponding to several
lepton momentum bins.
The net number of tags with second leptons above 1.5~GeV/c is
$1268\pm 54$ (statistical error only).

The net number of tags in each bin of second lepton momentum
is now corrected for the lepton identification efficiency.
We use the combined efficiency for geometric acceptance, track 
reconstruction, and identification as a function of momentum
which is found in Ref.~\cite{rwang}.

We now estimate and subtract the contribution from $\psi$ decay.
The spectrum of electrons or muons from
$\psi$ decay which accompany selected tags 
is found using Monte Carlo $B\bar B$ events.
It is then scaled to the number of $\psi$'s in data.
The scaling factor is determined through reconstruction of decays
$\psi\rightarrow\mu^+\mu^-$ in the full event set and in Monte Carlo.

\begin{figure}
\centerline{\psfig{figure=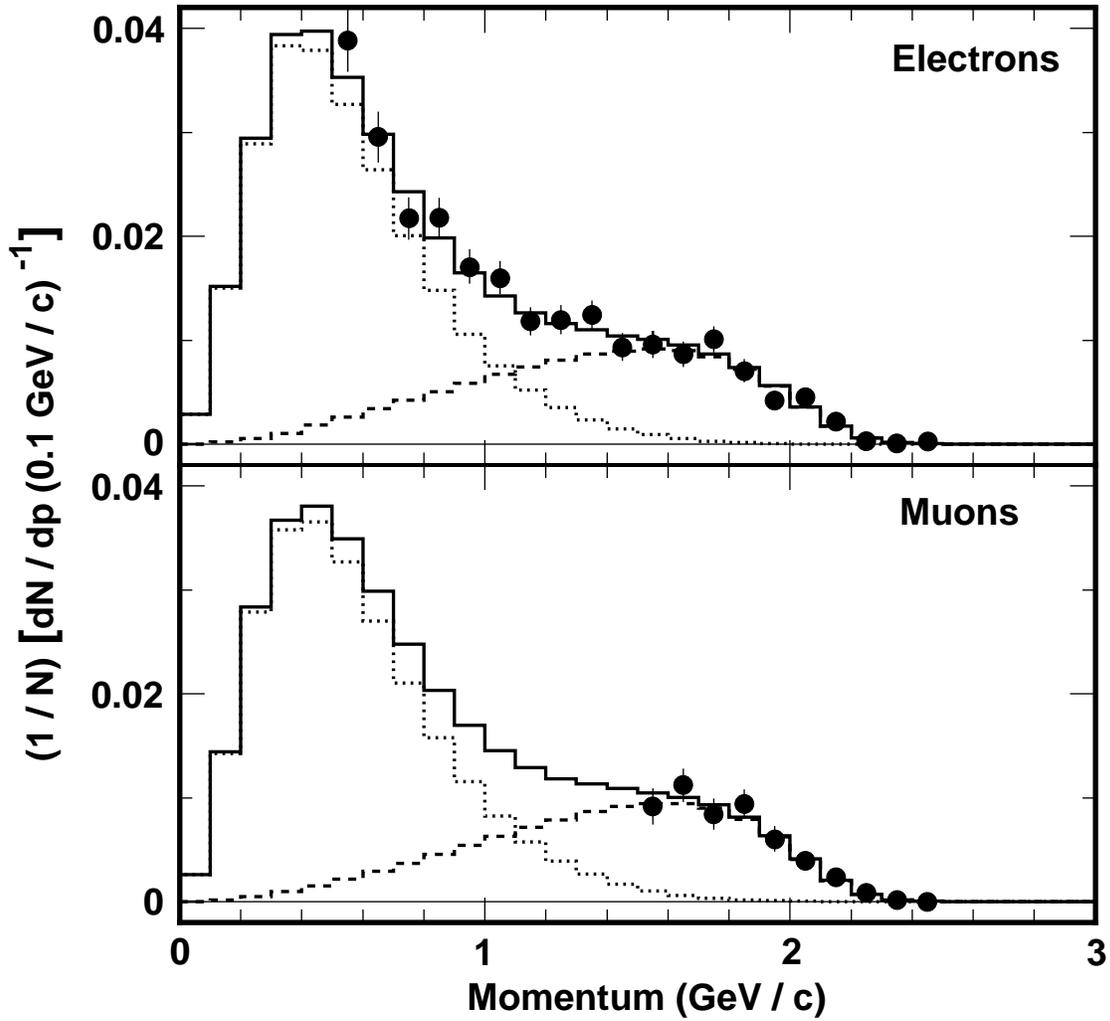}}
 \caption{Measured spectrum of second leptons, corrected for efficiency
and with $\psi$ contribution subtracted, shown with fit to
primary (dashed) and secondary (dotted) spectra.}
 \label{fig:spec_fit}
\end{figure}

The remaining second leptons can be considered as either primary,
produced directly in semileptonic $B$ decay, or secondary,
from products of $B$ decay.
To assess the contribution from secondary decay,
the tagged electron and muon spectra are now fitted
simultaneously to sums of primary and
secondary shapes, as was done for the inclusive spectra \cite{rwang}.
The primary shape is derived using the model of Altarelli 
{\it et al.} \cite{lepmodel}.
Secondary leptons originate from
decays of charm ($D$, $D_s$, $\Lambda_c$) and $\tau$, as
well as (for electrons) $\pi^0$ Dalitz decays and conversions of photons in the
beam pipe.
Above 0.6~GeV/c, the spectra from all of these are similar in shape.
They are therefore approximated using the spectrum from $D$'s.
The inclusive spectrum of Ref.~\cite{rwang}, which was explicitly
corrected for all but the $D$ contribution, was fitted using
the same shape.
The tagged spectra, normalized to $N_T$, are shown with fits in 
Figure~\ref{fig:spec_fit}.
Their overall appearance 
differs from that of the  inclusive spectra because the secondary 
rate is higher relative to the primary rate; 
as one primary lepton is in the tag, there is at most one other
primary lepton, whereas there can be two secondaries from charm.

The secondary contribution is now subtracted from both tagged
and inclusive spectra.
We then apply two corrections to the tagged spectrum.
The first, arising from the efficiency of
requirements on $\cos\theta_{\ell\ell}$ and on track 
multiplicity, has a net value of $+1.1\pm 1.1\%$ \cite{mssthesis}.
The second  arises because the tagging efficiency has 
a slight dependence on the decay charge multiplicity of
the other $B$ in an event, such that if the other decay is semileptonic,
its lower average multiplicity results in a higher tag efficiency.
The correction was calculated for each momentum bin of the second
lepton using a Monte Carlo simulation and found to be 
independent of second lepton momentum, with an average value of
$(-2.9\pm 1.3)$\%.
The net correction to the tagged rate from the two
sources is $(-1.8\pm 1.7)$\%.

\begin{figure}
\centerline{\psfig{figure=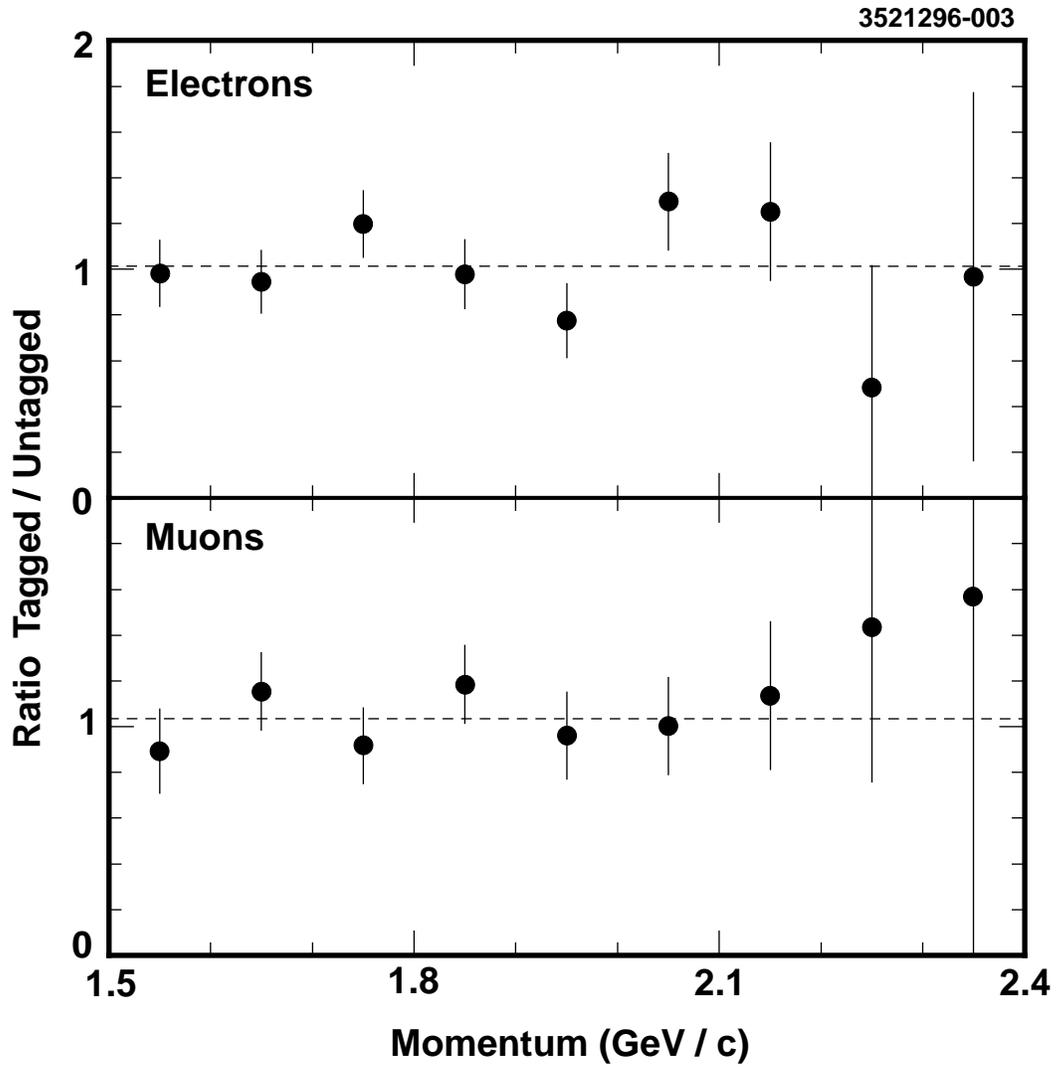}}
 \caption{Bin-by-bin ratio of tagged and inclusive spectra, 
with fits to a constant (dashed line) 
for
electrons (top) and muons (bottom).
}
 \label{fig:rwratio}
\end{figure}

The tagged spectra of electrons and muons are 
then divided by the corresponding
inclusive spectra, bin-by-bin.
If the primary spectra have the same shape, the resulting
distribution should be flat, with a value equal to
$R\equiv {b_{tag}/b_{incl}}$, where $b_{tag}$
and $b_{incl}$ are the semileptonic branching fractions measured in
the tagged and total events samples, respectively.
The distributions appear indeed to be flat.
They are thus fitted to a constant function, where we now restrict 
the momentum range to 1.5-2.4~GeV/c in order to limit
uncertainties associated with
the secondary spectrum (Figure~\ref{fig:rwratio}).
The fits yield $R=1.013\pm 0.061\ (1.036\pm 0.071)$ with $\chi^2/DOF=7.5/8$
(3.1/8) for electrons (muons).
The weighted average is 
\begin{eqnarray*}
R=1.023\pm 0.046.
\end{eqnarray*}
We also obtain $R$ by taking the ratio of
the primary contributions to the spectra from the fits described
above, and find excellent agreement.

%
Because the procedures for obtaining tagged and inclusive spectra
have been made nearly identical, many of the associated
systematic
errors are highly correlated and cancel approximately in the ratio $R$.
Among these are errors from lepton tracking
and identification efficiencies, subtraction of the $\psi$
contribution, fake rates, and spectral shapes for fitting
to find the secondary contribution.
To estimate the effects of variations in the $\psi$ contribution
and fake efficiencies, we repeat our measurement,
adjusting each contribution in turn to the limits of the estimated
uncertainty.
To assess the uncertainty due to shapes used in fitting to obtain the
secondary contribution, fitting was repeated 
with varying primary and secondary shapes.
For the primary shape we include
the model of Isgur {\it et al.} \cite{isgur}
where the fractional contribution of $B\rightarrow D^{**}l\nu$ 
to semileptonic decays is fixed for each fit in the range 
$(23\pm 1)$\%\cite{rwang} (ISGW$^{**}$).
For the secondary spectrum we use the method and shapes used in
Ref.~\cite{mix}.
These correlated sources and effects are summarized in Table~\ref{tab:syserr1}.

Uncertainties which apply to the tagged but not to the inclusive
measurement 
include effects due to event selection, angular cuts,
and tagging efficiency differences discussed above, which contribute
a total of 1.7\%.
The uncertainty from the background 
estimate for the tag cancels approximately
in dividing by $N_T$, and we assign an overall
error of 1\% to the magnitude of the tagged spectrum from this source.
In addition, we approximate the spectra of all secondary leptons in the tagged
spectra by those from $D$ decay only.
To estimate the uncertainty from this and from the fact that
there may be small differences in shape between
tagged and inclusive samples due to tag selection
itself, we repeat the analysis after subtracting an
estimated contribution of secondaries from sources other than $D$ decay.
This causes a shift of 1.7\%, which is taken to be the error.
The net systematic error on the value of $R$ is 3.5\%.

$R$ may be represented by an equation in terms of
$b_0$, $b_+$, $\beta\equiv{b_+/ b_0}$, and several other
parameters, which can be solved for $b_0$, $b_+$, and $\beta$.
The average semileptonic branching fractions of $B$'s in the tagged and
inclusive event samples,
$b_{tag}$ and $b_{incl}$,
may be expressed as  
\begin{eqnarray}
b_{incl} & = & f_{00}b_0 + f_{+-}b_+=f_{00}b_0(1+\alpha) \label{eq:bincl}\\
b_{tag} & = & f_{0,tag}b_0 + f_{+,tag}b_+=b_0(1+f_{+,tag}\delta\beta), \label{eq:btag}
\end{eqnarray}
where
$f_{00}$ and $f_{+-}$ are the fractions of $B^0$ and $B^+$ events
from the $\Upsilon$(4S),
$f_{0,tag}$ and $f_{+,tag}$ are the corresponding fractions
among tags, 
$\alpha \equiv {f_{+-}b_+/ f_{00}b_0}$, and 
$\delta\beta\equiv\beta-1$.

From (\ref{eq:btag}) and the definition of $\beta$ we obtain
\begin{eqnarray}
b_0 & = & {b_{tag}\over 1 + f_{+,tag}\delta\beta}
= {R\ b_{incl}\over 1 + f_{+,tag}\delta\beta} \label{eq:b0} \\
b_+ & = & \beta b_0 
 =  {R\ b_{incl}(1+\delta\beta)\over 1 + f_{+,tag}\delta\beta}
\label{eq:b+}
\end{eqnarray}
To obtain $\delta\beta$, we take the ratio of (\ref{eq:btag}) and (\ref{eq:bincl})
\begin{eqnarray}
R& \equiv & {b_{tag}\over b_{incl}}=
{b_0(1+f_{+,tag}\delta\beta)\over f_{00}b_0(1+\alpha)},\\
\Rightarrow f_{00}&=&{1+f_{+,tag}\delta\beta\over R(1+\alpha)}. \label{eq:f001}
\end{eqnarray}
From the definitions of $\alpha$ and $\delta\beta$,
\begin{eqnarray}
f_{00}& = & {1+\delta\beta \over \alpha + 1 + \delta\beta}. \label{eq:f002}
\end{eqnarray}
Equating (\ref{eq:f001}) and (\ref{eq:f002}),
\begin{eqnarray}
(1+\delta R)(1+\alpha)(1+\delta\beta)&=& (1+f_{+,tag}\delta\beta)(\alpha + 1+\delta\beta) \label{eq:quad}
\end{eqnarray}
$f_{+,tag}$, $\delta\beta$, and 
$\delta R \equiv R-1$ are expected to be ``small'', $< 0.1$,
so we discard the highest order term, 
$f_{+,tag}\delta\beta^2$, to
arrive at a linear equation in $\delta\beta$ which
is solved to obtain
\begin{eqnarray}
\delta\beta = {\delta R(1+\alpha)\over (1+\alpha)(f_{+,tag}-\delta R) -\alpha}.
\end{eqnarray}

For $\alpha$ we average the CLEO~II measurements $\alpha = {{
{\cal B}{(\Upsilon(4S)\rightarrow B^+B^-,B^+\rightarrow \bar D^{*0}\ell^+\nu})
/
{\cal B}{(\Upsilon(4S)\rightarrow B^0\bar B^0, B^0\rightarrow D^{*-}\ell^+\nu})}
=1.14\pm 0.14\pm 0.13}$\cite{alpha}
and $\alpha = {{
{\cal B}{(\Upsilon(4S)\rightarrow B^+B^-,B^+\rightarrow \psi K^{(*)+})}
/
{\cal B}{(\Upsilon(4S)\rightarrow B^0\bar B^0, B^0\rightarrow\psi  
K^{(*)0})}}=1.15\pm 0.17\pm 0.06
}$\cite{psik},
to get $\alpha = 1.15\pm 0.13$.

The fraction $f_{+,tag}$ is nonzero because
some charged $B$'s are present among our tags from decays of
the type $B\rightarrow D^{*-}\pi\ell^+\nu$ (where the
$D^{*+}\pi$ may or may not form a resonance).
The fraction $f^{**}$ of such decays in the tag sample may be determined
by adding the reconstruction of the $\bar D^0$ decay,
so that a measured $D^{*-}$ momentum is used in the calculation of 
missing mass squared.
In this case the resolution is sufficient to achieve some degree of
separation between the decays without and with an additional pion,
so that the relative rates may be determined.
In terms of our analysis the only directly applicable measurement has been
performed at CLEO~I, with the same event and tag lepton momentum requirements 
which found  $f^{**}=0.14\pm 0.08$, the value of which can
be derived from measurements cited in \cite{daniela}.
Because $f^{**}$ is a ratio, it depends mainly on kinematic criteria
and should not differ significantly in CLEO~II.
We use a measurement from ALEPH,
${\cal B}(B\rightarrow D^{*-}\pi^+\ell^+\nu X)=
(1.25\pm 0.16\pm 0.12)\%$ \cite{lepd2st},
and Monte Carlo simulations to obtain an independent value, 
$f^{**}=0.10\pm 0.02$, which is in agreement with CLEO~I but
may have a large systematic error due to insufficiency of our
current knowledge about the different modes which contribute
to $B\rightarrow D^{*-}\pi^+\ell^+\nu X$.
We use the CLEO~I value because it is more directly obtained,
and  our evaluations of $b_0$, $b_+$, and $f_{00}$ are 
insensitive to it.
The relative contributions to $f^{**}$ from charged and 
neutral $B$'s is determined using the approximate
conservation of isospin in the fragmentation of the hadronic part
of the final state;
if $b_0$ and $b_+$ are equal and the charged and neutral $B$'s decay to
$\bar D^*X\ell^+\nu$ at the same rates, the rate to states where 
$X$ is a charged pion is expected to be
double that of the rate where it is a neutral pion.
In other words, $B^+\rightarrow D^{*-}\pi^+\ell^+\nu$ contributes 
to $f^{**}$ at twice the rate of 
$B^0\rightarrow D^{*-}\pi^0\ell^+\nu$,
weighted by the respective production rates and semileptonic
branching fractions \cite{mix},
\begin{eqnarray}
f_{+,tag} & = & f^{**}{2\alpha\over 1+2\alpha}, \label{eq:fptag}
\end{eqnarray}
from which we get $f_{+,tag}=(9.7\pm 5.5)\%$.

Using 
$R = 1.023\pm 0.046\pm 0.036$,
$f^{**} = 0.14\pm 0.08$, and
$\alpha = 1.15\pm 0.13$, 
we obtain
\begin{eqnarray*}
\beta & = & 0.950 {^{+0.117}_{-0.080}}{^{+0.091}_{-0.068}}, \\
f_{+-}/f_{00} & = & 1.21 \pm 0.12 \pm 0.17,
\end{eqnarray*}
where in each case the first error is statistical and the
second systematic.
The latter number may be expressed 
$f_{00} =  0.452 \pm 0.028 \pm 0.038$.
The values of $b_0$ and $b_+$ may be obtained with the
addition of $b_{incl}$.
The most recent published measurement of an inclusive semileptonic
branching fraction at the $\Upsilon$(4S) was made using a 
lepton tagging method which gives a value that is not strictly 
equal to $b_{incl}$\cite{dilep}, as the tagged event sample may
not have the same composition as generic $\Upsilon$(4S) events.
However, given known limits on the input parameters, it is expected to
differ from $b_{incl}$ by less than one part in $10^2$, so we use
the result, $b_{incl} \approx (10.49\pm 0.46)\%$, to get
\begin{eqnarray*}
b_0 & = & (10.78 \pm 0.60 \pm 0.69)\%,\\
b_+ & = & (10.25 \pm 0.57 \pm 0.65)\%. 
\end{eqnarray*}
As the determinations of $b_0$, $b_+$, $\beta$, and $f_{+-}/f_{00}$ 
involve correlated inputs,
the errors are obtained numerically.
All of the statistical uncertainties are defined by the 
uncertainty in $R$.
The dominant contributors to the systematic uncertainties are 
$R$ for $\beta$, both $R$ and $b_{incl}$ for $b_0$ and $b_+$, 
and $\alpha$ for $f_{00}$.

Our results are consistent with our previous ones \cite{b0},
$\beta = 0.93 \pm 0.18 \pm 0.12$,
$b_0 = (10.9 \pm 0.7\pm 1.1)$\%, and
$b_+ = (10.1 \pm 1.8\pm 1.5)$\%. 
Our new result for $b_0$ is somewhat correlated with the previous one,
as the data sets overlap partially and parts of the method are
similar.
The results for $b_+$ are uncorrelated.
$f_{+-}/f_{00}$ may also be obtained using the CLEO measurement 
of $\alpha$ with a
lifetime ratio based on measurements at LEP and CDF, as in \cite{alpha}.
The value reported here was obtained using CLEO 
data exclusively.

To summarize,
the electronic and muonic spectra have been measured in a sample of events
enriched in neutral $B$'s using partial reconstruction tags.
By comparing these spectra to the inclusive spectra in
$\Upsilon$(4S) events, 
we probe differences between $B^0$ and $B^+$
in rates  and spectral shapes of primary semileptonic decay. 
We observe no evidence for differences in the 
spectra between 1.5 and 2.4~GeV/c.
Assuming the shapes are the same, we obtain
$b_0$, $b_+$, $\beta$, and $f_{+-}/f_{00}$.
Our new value of $f_{+-}/f_{00}$ is the first that is independent
of $B$ meson lifetime measurements.

\centerline{\bf ACKNOWLEDGEMENTS}
\smallskip
We gratefully acknowledge the effort of the CESR staff in providing us with
excellent luminosity and running conditions.
J.P.A., J.R.P., and I.P.J.S. thank                                           
the NYI program of the NSF, 
M.S. thanks the PFF program of the NSF,
G.E. thanks the Heisenberg Foundation, 
K.K.G., M.S., H.N.N., T.S., and H.Y. thank the
OJI program of DOE, 
J.R.P., K.H., M.S. and V.S. thank the A.P. Sloan Foundation,
and A.W. and R.W. thank the 
Alexander von Humboldt Stiftung
for support.
M.S. is supported as a Cottrell Scholar of Research Corporation.
This work was supported by the National Science Foundation, the
U.S. Department of Energy, and the Natural Sciences and Engineering Research 
Council of Canada.

%
%

\begin{table}
\begin{center}
\caption{Systematic errors on $R$.
Those listed in the upper section apply both to tagged and 
inclusive spectra,
while those in the lower section apply only to the tagged spectra.
For each listed source, the ratio of its magnitude to the net
tagged spectrum above 1.5~GeV/c is shown in parentheses.}
\begin{tabular} {ccc}
Source & ${\sigma_{sys}/{value}}(\%)$ 
& ${\delta(R)/ R}$(\%) \\
& or procedure & \\
\hline
$\psi$ contribution (0.011)     & 30         & 0.3 \\
Fake electrons  (0.002) & 50         & 0.1\\
Fake muons  (0.011) &25          & 0.2 \\
Secondary spectrum (.044)& as in Ref.~\cite{mix}   & 2.3 \\
Primary spectrum (0.95) & ISGW**\cite{rwang}         & $0.3$ \\
$(b\rightarrow u)$ (0.019)& 21     & -- \\ \hline
Event/tag selection &           1.7        & 1.7 \\
Background estimate (0.59) &           1.0        & 1.0 \\
Secondary spectrum (0.044) &   vary sources (see text)        & 1.7 \\ \hline
Net systematic error	&		&     3.5
\end{tabular}
\label{tab:syserr1}
\end{center}
\end{table}

\end{document}